\documentstyle[twoside,fleqn,espcrc2,epsf]{article}

\newcommand{\AmS}{{\protect\the\textfont2
  A\kern-.1667em\lower.5ex\hbox{M}\kern-.125emS}}

\title{On weak coupling expansion in models with
unbroken symmetry
\thanks{Work supported by Bundesministerium f\"ur Wissenschaft,
und Transport of Austria}
        \thanks{e-mail:oleg@ap3.bitp.kiev.ua}}

\author{O. Borisenko, \address{Institute for Theoretical Physics, 
National Academy of Sciences of Ukraine, Kiev 252143, Ukraine}}

\begin{document}

\begin{abstract}
An investigation of the weak coupling region of $2D$ $SU(N)$ 
spin models is presented. An expansion of the free energy 
and correlation functions at low temperatures is performed 
in the link formulation with periodic boundary conditions (BC). 
The resulting asymptotics is shown to be nonuniform in the volume
for the free energy. 
\end{abstract}


\maketitle

This paper deals with $2D$ $SU(N)\times SU(N)$ principal chiral model in the
weak coupling region whose partition function (PF) is given by
\begin{equation}
Z = \int\prod_xDU_x
\exp \left [\beta \sum_{x,n} {\mbox{ Re Tr}} U_xU_{x+n}^+\right] , 
\label{pfsun}
\end{equation}
\noindent
where $U_x\in SU(N)$, $DU_x$ is the invariant measure and we
assume periodic BC. 
The main question discussed here is what is the correct asymptotic
expansion of nonabelian models at large $\beta$ and whether the conventional
perturbation theory (PT) gives the correct expansion, the question
addressed few years ago in \cite{superinst}. As was rigorously proven,
the conventional PT gives an asymptotic expansion which is uniform
in the volume for the abelian $XY$ model \cite{XYPT}. One of the basic
theorem which underlies the proof states that 
the following inequality holds in the $3D$ $XY$ model
\begin{equation}
< \exp (\sqrt{\beta}A(\phi_x)) > \ \leq \ C ,
\label{ineq3D}
\end{equation}
\noindent
where $C$ is $\beta$-independent and $A(\phi +2\pi)=\phi$. $\phi_x$
is an angle parametrizing the action of the $XY$ model
$S=\sum_{x,n}\cos (\phi_x - \phi_{x+n})$. It follows, at large
$\beta$ the Gibbs measure is concentrated around $\phi_x\approx 0$ 
giving a possibility to construct an expansion around $\phi_x=0$. 
This inequality is not true in $2D$ because of the Mermin-Wagner 
theorem (MWT), however the authors of \cite{XYPT} prove the same
inequality for the link angle, i.e. 
\begin{equation}
< \exp (\sqrt{\beta}A(\phi_l)) > \ \leq \ C \ , \phi_l=\phi_x-\phi_{x+n}.
\label{ineq2D}
\end{equation}
\noindent
Thus, in $2D$ the Gibbs measure at large $\beta$ is concentrated
around $\phi_l\approx 0$ and the asymptotics can be
constructed expanding in powers of $\phi_l$. In the abelian case such
an expansion is equivalent to the expansion around $\phi_x=0$
because i) the action depends only on the difference $\phi_x - \phi_{x+n}$
and ii) the integration measure is flat, $DU_x=d\phi_x$. 

I'm not aware of any similar rigorous result proven for the nonabelian
model. Nevertheless, because of the MWT one has to expect something
similar to (\ref{ineq2D}), namely
\begin{equation}
< \exp (\sqrt{\beta}{\mbox{arg}}A({\mbox{Tr}}V_l)) > \leq C \ ,
V_l=U_xU_{x+n}^+.
\label{nabineq2D}
\end{equation}
\noindent
This is equivalent to the statement that the Gibbs measure at large
$\beta$ concentrates around $V_l\approx I$. That such (or similar)
inequality holds in $2D$ models is intuitively clear and should also
follow from chessboard estimates. It is assumed here that 
(\ref{nabineq2D}) is correct, hence $V_l=I$ is the only saddle point
for the invariant integrals. Thus, correct asymptotic expansion
should be given via expansion around $V_l=I$, similarly to the
abelian model. If the conventional PT gives correct expansion, it
must reproduce the series obtained by expansion around $V_l=I$.
However, neither i) nor ii) holds in the nonabelian models, therefore
it is far from obvious that two expansions indeed coincide. 
Our purpose here is to develop an expansion around $V_l=I$ 
aiming to calculate the asymptotic series for nonabelian models. 

To accomplish this task we make a change of variables
$V_l=U_xU_{x+n}^+$ in (\ref{pfsun}). PF becomes \cite{linkrepr}
\begin{equation}
Z = \int \prod_l dV_l 
\exp \left[ \beta \sum_l {\mbox{Re Tr}} V_l + \ln J(V) \right] ,
\label{lPF}
\end{equation}
\noindent
\begin{equation}
J(V) = \prod_p 
\left[ \sum_r d_r \chi_r \left( \prod_{l\in p}V_l \right) \right].
\label{jacob}
\end{equation}
\noindent
$\prod_p$ is a product over all plaquettes of $2D$ lattice,
the sum over $r$ is sum over all representations of $SU(N)$, 
$d_r=\chi_r(I)$ and
\begin{equation}
\prod_{l\in p}V_l = V_n(x)V_m(x+n)V_n^+(x+m)V_m^+(x).
\label{prod}
\end{equation}
\noindent
$J(V)$ is a product of $SU(N)$ delta-functions 
which introduce the constraint $\prod_{l\in p}V_l=I$
\footnote{On the periodic lattice one has to constraint two 
holonomy operators, i.e. closed paths winding around the whole lattice. 
Such global constraints do not influence thermodynamic limit in $2D$.}. 
Its solution is a pure gauge $V_l=U_xU_{x+n}^+$, 
thus two forms of the PF are equivalent. 
The correlation function $\Gamma (x,y)=< {\mbox{Tr}} U_xU_y^+ >$ becomes
\begin{equation}
\Gamma (x,y) = < {\mbox{Tr}} \prod_{l\in C_{xy}} W_l > \ ,
\label{corf1}
\end{equation}
\noindent
where $C_{xy}$ is some path connecting points $x$ and $y$ and 
$W_l = V_l$ if along the path $C_{xy}$ the link $l$ goes in
the positive direction and $W_l = V_l^+$, otherwise.  
The abelian analog of (\ref{lPF}) reads
\begin{equation}
Z_{XY} = \int\prod_l\left [d\phi_le^{\beta \cos\phi_l}\right ]
\prod_p\sum_{r=-\infty}^{\infty} e^{ir\phi_p}, 
\label{PFLxy}
\end{equation}
\noindent
$\phi_p=\phi_n(x)+\phi_m(x+n)-\phi_n(x+m)-\phi_m(x+n)$.

The main step is to find a form for (\ref{lPF})
appropriate for large-$\beta$ expansion. We consider 
$SU(2)$ model and parametrize ($\sigma^k$ are Pauli matrices)
\begin{equation}
V_l = \exp [i\sigma^k\omega_k(l)] \ , \ k=1,2,3 \ ,
\label{su2mtr}
\end{equation}
\noindent
\begin{equation}
W_l = \left[ \sum_k\omega^2_k(l) \right]^{1/2} , \
W_p = \left[ \sum_k\omega^2_k(p) \right]^{1/2} ,
\label{Wlp}
\end{equation}
\noindent
where $\omega_k(p)$ is a plaquette angle defined as
\begin{equation}
V_p = \prod_{l\in p}V_l =  \exp [i\sigma^k\omega_k(p)] .
\label{plangle}
\end{equation}
\noindent
Then, the PF (\ref{lPF}) can be exactly rewritten to the following 
form on a dual lattice ($p\to x$)
\begin{eqnarray}
Z=\int \prod_l \left[ \frac{\sin^2W_l}{W^2_l}e^{\beta \cos W_l}
\prod_kd\omega_k(l) \right] \nonumber     \\
\prod_x \frac{W_x}{\sin W_x} \prod_x \sum_{m(x)=-\infty}^{\infty}
\int\prod_kd\alpha_k(x)  \nonumber     \\
\exp \left[ -i\sum_k\alpha_k(x)\omega_k(x) + 2\pi im(x)\alpha (x) \right] \ ,
\label{PFwk}
\end{eqnarray}
\noindent
where we have introduced the auxiliary field $\alpha_k(x)$ and
$\alpha (x) = (\sum_k\alpha^2_k(x))^{1/2}$ .
The derivation of the PF (\ref{PFwk}) is given in \cite{asexp}.
To perform large-$\beta$ expansion we proceed in a standard way, 
i.e. first we make the substitution
\begin{equation}
\omega_k(l)\to (\beta)^{-1/2}\omega_k(l) \ , \
\alpha_k(x)\to (\beta)^{1/2}\alpha_k(x) 
\label{subst}
\end{equation}
\noindent
and then expand in powers of fluctuations of link fields.
Technical details are left to a forthcoming 
paper \cite{asexp}. The essential aspects are:

1). The main building blocks of the expansion are
the following functions which we term ``link'' Green functions
\begin{eqnarray}
G_{ll^{\prime}} = 2\delta_{l,l^{\prime}} - G_{x,x^{\prime}} -
G_{x+n,x^{\prime}+n^{\prime}} + G_{x,x^{\prime}+n^{\prime}} 
\nonumber  \\
+ G_{x+n,x^{\prime}},  \ D_l(x^{\prime}) = G_{x,x^{\prime}} -
G_{x+n,x^{\prime}}.
\label{Gll1}
\end{eqnarray}
\noindent
$G_{x,x^{\prime}}$ is a ``standard'' Green function on the periodic
lattice. Unlike $G_{x,x^{\prime}}$, both $G_{ll^{\prime}}$ 
and $D_l(x^{\prime})$ are infrared finite by construction.

2). Sums over representations are treated via the Poisson resummation
formula. Due to this, zero modes decouple from the expansion 
after integration over the auxiliary field. 

3). An ensemble of fluctuations around $V_l=I$ can be 
written as usual Gaussian ensemble appearing due to integration 
over representations (over auxiliary field) in abelian  
(nonabelian) case.

4). Generating functionals (GF) appear to be:
\begin{equation}
M(h_l) = \exp \left [ \ \frac{1}{4} 
h_lG_{ll^{\prime}}h_{l^{\prime}} \ \right ], \ G_{ll}=1
\label{Gfunxy1}
\end{equation}
\noindent
for the $XY$ model and
\begin{eqnarray}
M(h_l,s_x) = \exp 
[ \ \frac{1}{4}s_k(x)G_{x,x^{\prime}}s_k(x^{\prime}) 
\nonumber   \\
- \frac{i}{2}s_k(x)D_l(x)h_k(l) + 
\frac{1}{4}h_k(l)G_{ll^{\prime}}h_k(l^{\prime})\ ] \ ,
\label{GFfin}
\end{eqnarray}
\noindent
for the $SU(2)$ model. $h_k(l)$ ($s_k(x)$) is a source for the link
(auxiliary) field and sum over all repeating indices
is understood. We list below some results proven in \cite{asexp}.

I. Asymptotics for the free energy of the $XY$ model
can be obtained from the expansion
\begin{equation}
Z_{XY}\sim\prod_{l} \left [ 1+\sum_{k=1}^{\infty}\frac{1}{\beta^k}
A_k ( \frac{\partial^2}{\partial h_l^2} ) \right ] M(h_l) \ ,
\label{asxyGll}
\end{equation}
\noindent
where $A_k$ are known coefficients. It is easy to prove that
this expansion coincides with the standard PT answer. For example,
the first order coefficient is given by
\begin{equation}
\frac{1}{L^2} C_{XY}^1 = 
\frac{1}{32 L^2} \sum_{l} G_{ll}^2 = \frac{1}{16} \ .
\label{C1Gll}
\end{equation}
\noindent

II. Correlation function (\ref{corf1}) in the $SU(2)$ model 
up to the first order is given by
\begin{equation}
\Gamma (x,y) = 1 - \frac{3}{4\beta} 
\sum_{l,l^{\prime}\in C^d_{xy}} G_{ll^{\prime}} + O(\beta^{-2}) \ ,
\label{GLXY1}
\end{equation}
\noindent
where $C^d_{xy}$ is a path dual to the path $C_{xy}$, i.e consisting
of the dual links which are orthogonal to the original
links $l,l^{\prime}\in C_{xy}$.
The form of $G_{ll^{\prime}}$ guarantees independence of $\Gamma (x,y)$
of the choice of the path $C_{xy}$. After some algebra it is easy to recover
the standard PT answer
\begin{eqnarray}
\Gamma (x,y) = 1 - \frac{3}{2\beta} D(x-y) \ ,   \\
D(x) = \frac{1}{L^2} \sum_{k_n=1}^{L-1}
\frac{1 - e^{\frac{2\pi i}{L}k_n x_n}}
{D-\sum_{n=1}^D\cos \frac{2\pi}{L}k_n} \ . \nonumber
\label{Dx}
\end{eqnarray}
\noindent

III. The main result of this study is the first order
coefficient of the $SU(2)$ free energy. The second order 
term in the auxiliary fields gives the following contribution
at this order
\begin{equation}
\frac{1}{L^2}C_{div}^1=\frac{3}{8L^2}\sum_{x,x^{\prime}}
G_{x,x^{\prime}}Q_{x,x^{\prime}} \ .
\label{Cdiv}
\end{equation}
\noindent
$Q_{x,x^{\prime}}$ is a second order polynomial in link Green
functions (\ref{Gll1}). Numerical study of (\ref{Cdiv})
shows that $\frac{1}{L^2}C_{div}^1=b+a_1\ln L + O(\frac{\ln L}{L})$.
All other contributions coming from the action, from the measure
and from $J(V)$ depend only on the link Green functions and
are finite. The final result is
\begin{equation}
\frac{1}{L^2}C_{SU(2)}^1 = a_0 + a_1\ln L 
+ O(\frac{\ln L}{L}) \ ,
\label{su2fr}
\end{equation}
\noindent
where $a_i$ are $\beta$- and $L$-independent.

{\bf Conclusions:}
The GF of the $XY$ model depends only on 
$G_{ll^{\prime}}$, guaranteeing the infrared finitness of 
the expansion. This is a direct consequence of the fact that the
action of $XY$ model is a function of $\phi_l$. It is not the
case in nonabelian models: the GF (\ref{GFfin})
includes also dependence on $G_{x,x^{\prime}}$ creating
potential danger for nonuniformity. The first order
coefficient of $\Gamma (x,y)$ is expressed only via 
$G_{ll^{\prime}}$ recovering thus the standard PT result. 
A genuine nonabelian contribution appears only in the first order 
coefficient of the free enegy or in the second order coefficient 
of $\Gamma (x,y)$. That there might be a problem with 2-nd order 
coefficient of $\Gamma (x,y)$ has been shown in \cite{superinst},
namely the PT with superinstanton BC (SIBC) gives a result different
from the PT result obtained with periodic or Dirichlet BC. 
As was shown later, the PT with SIBC
diverges at the third order \cite{sup2lp}. The interpretation
has been given that it is just SIBC for which
the asymptotics is nonuniform in the volume. Our result
seems to support other interpretation: 
the correct asymptotic expansion of $2D$ nonabelian 
models is nonuniform in the volume. In particular, it means
that the conventional PT does not take into account all the 
saddle points when the volume is increasing and becomes larger
than the perturbative correlation length.
We conclude thus either 1) to get the correct expansion one has to
expand a true function in the thermodynamic limit (like in $1D$)
or 2) the inequality (\ref{nabineq2D}) does not hold, e.g.
the factor $\sqrt{\beta}$ is incorrect or 3) nonabelian models
are not expandable in $1/\beta$ in two dimensions.


\begin{thebibliography}{99}

\bibitem{superinst} A.~Patrascioiu, E.~Seiler, Phys.Rev.Lett.
74 (1995) 1920; Phys.Rev. D57 (1998) 1394.
%
\bibitem{XYPT} J.~Bricmont, J.-R.~Fontaine, J.L.~Lebowitz, E.H.~Lieb, 
T.~Spencer, Comm.Math.Phys. 78 (1981) 545.
%
\bibitem{linkrepr} B.~Rusakov, Phys.Lett. B398 (1997) 331.
%
\bibitem{asexp} O.~Borisenko, V.~Kushnir, V.~Velytsky, 
to appear in hep-lat.
%
\bibitem{sup2lp} F.~David, Phys.Rev.Lett. 75 (1995) 2626;
F.~Niedermayer, M.~Niedermaier, P.~Weisz, hep-lat/9612002.
%


\end{thebibliography}
\end{document}